\documentclass[runningheads]{llncs}
\usepackage{graphicx}
\usepackage[colorlinks = true,
            linkcolor = blue,
            urlcolor  = blue,
            citecolor = blue,
            anchorcolor = blue]{hyperref}
\usepackage{framed}

\begin{document}

%
%\title{Contribution Title\thanks{Supported by organization x.}}
%\title{SmmPack: Towards a Framework for Obfuscating SMM Modules Against Attacker Vulnerability Analysis}

\title{SmmPack: Obfuscation for SMM Modules with TPM Sealed Key}

\newenvironment{mybullet}{\begin{list}{$\bullet$}
  {\setlength{\topsep}{0mm}\setlength{\itemsep}{0mm}
    \setlength{\parsep}{0.5mm}
    \setlength{\itemindent}{0mm}\setlength{\partopsep}{0mm}
    \setlength{\labelwidth}{15mm}
    \setlength{\leftmargin}{5mm}}}{\end{list}}

\titlerunning{SmmPack: Obfuscation for SMM Modules}
% If the paper title is too long for the running head, you can set
% an abbreviated paper title here

\author{
\begin{framed}
\normalsize
\raggedright
If you cite this paper, please use the following reference:\\Kazuki Matsuo, Satoshi Tanda, Yuhei Kawakoya, Kuniyasu Suzaki, and Tatsuya Mori, ``SmmPack: Obfuscation for SMM Modules,'' {\it Proceedings of the 21st Conference on Detection of Intrusions and Malware and Vulnerability Assessment} (DIMVA 2024), July 2024
\end{framed}
\leavevmode
Kazuki Matsuo\inst{1}\orcidID{0009-0002-0402-5312} \and
Satoshi Tanda\inst{4} \and
Kuniyasu Suzaki\inst{2}\orcidID{0000-0003-0912-0087} \and
Yuhei Kawakoya\inst{3}\orcidID{0009-0005-9310-0493} \and
Tatsuya Mori\inst{1,5,6}\orcidID{0000-0003-1583-4174}}

\authorrunning{K. Matsuo et al.}
% First names are abbreviated in the running head.
% If there are more than two authors, 'et al.' is used.

\institute{Waseda University, 1-104 Totsukamachi, Shinjuku-ku, Tokyo 169-8050, Japan \and
Institute of Information Security., 2-14-1 Tsuruyacho, Kanagawa-ku, Yokohama-shi, Kanagawa 221-0835, Japan \and
% \email{lncs@springer.com}\\
% \url{http://www.springer.com/gp/computer-science/lncs} \and
NTT Security (Japan) KK, 
4-14-1 Sotokanda, Chiyoda-ku, Tokyo 101-0021, Japan \and
Satoshi’s System Programming Lab \\
\url{https://tandasat.github.io} \and
NICT, 4-2-1 Nuki-kitamachi, Kokubunji-shi, Tokyo 184-0015, Japan \and
RIKEN API, 15th Floor, Nihonbashi 1-4-1, Nihonbashi, Chuo-ku, Tokyo 103-0027, Japan}

% \email{\{abc,lncs\}@uni-heidelberg.de}}
%
\maketitle              % typeset the header of the contribution
\begin{abstract}
%The abstract should briefly summarize the contents of the paper in 15--250 words.
System Management Mode (SMM) is the highest-privileged operating mode of x86 and x86-64 processors. Through SMM exploitation, attackers can tamper with the Unified Extensible Firmware Interface (UEFI) firmware, disabling the security mechanisms implemented by the operating system and hypervisor. Vulnerabilities enabling SMM code execution are often reported as Common Vulnerabilities and Exposures (CVEs); however, no security mechanisms currently exist to prevent attackers from analyzing those vulnerabilities.
To increase the cost of vulnerability analysis of SMM modules, we introduced SmmPack. The core concept of SmmPack involves encrypting an SMM module with the key securely stored in a Trusted Platform Module (TPM). We assessed the effectiveness of SmmPack in preventing attackers from obtaining and analyzing SMM modules using various acquisition methods. Our results show that SmmPack significantly increases the cost by narrowing down the means of module acquisition.
Furthermore, we demonstrated that SmmPack operates without compromising the performance of the original SMM modules. We also clarified the management and adoption methods of SmmPack, as well as the procedure for applying BIOS updates, and demonstrated that the implementation of SmmPack is realistic.

%\keywords{First keyword  \and Second keyword \and Another keyword.}
\end{abstract}
\section{Introduction}
% 背景（SMMとは、SMMに攻撃するメリット、SMMモジュールのCVEの増加）
System Management Mode (SMM) is the highest-privileged operating mode in x86 and x86-64 processors. The SMM module, a type of Unified Extensible Firmware Interface (UEFI) module, operates in a memory area called the System Management RAM (SMRAM), which is accessible only during SMM. Attackers can arbitrarily modify the BIOS image and insert malware or bypass security mechanisms, such as Virtualization Based Security (VBS)~\cite{vbsbypass,vbsbypass2}, by escalating privileges to SMM. Consequently, vulnerability researchers are actively exploring SMM modules for exploits, as shown by rising CVE reports~\cite{increasingcves}.

% 背景（CVEは実装ミスが原因、既存のセキュリティ機構では防げないので新しく脆弱性解析を妨害する仕組みが必要、脅威モデルの定義）
Many vulnerabilities in CVEs are caused by implementation errors, such as memory access outside the SMRAM~\cite{smmcallout,smmcallout2,smm6cves}. To exploit these vulnerabilities, attackers first need to analyze the SMM module. 
Unfortunately, there are currently no security mechanisms that hinder attacker's vulnerability analysis. Consequently, developers of SMM modules are forced to implement these modules without vulnerabilities to defend against the exploits. Existing security mechanisms, such as secure boot, cannot prevent such attacks because attacks utilizing these vulnerabilities do not require modification to the UEFI firmware.
% Therefore, security mechanisms that increase the cost for attackers are necessary. In this study, we define the threat model as "attackers aiming to discover vulnerabilities in SMM modules."
Therefore, security mechanisms that increase the cost for attackers are necessary.
%, especially considering the threat we target in this study: 'attackers aiming to discover vulnerabilities in SMM modules.'

% \mori{
% Unfortunately, there are currently no security mechanisms that prevent an attacker from analyzing vulnerabilities. As a result, developers of SMM modules are forced to implement these modules without vulnerabilities to counter attacks that exploit them.
% However, existing mechanisms such as secure boot are not sufficient to prevent such attacks because these vulnerabilities can be exploited without requiring any changes to the UEFI firmware.
% Therefore, the development and implementation of security mechanisms specifically designed to raise the cost of vulnerability analysis for attackers is not only a viable strategy, but an indispensable one for improving security.
% }

%提案手法
We propose SmmPack as an obfuscation system to increase the cost of vulnerability analysis of SMM modules to attackers.
To the best of our knowledge, this is the first study to cautiously explore the application of obfuscation techniques to UEFI firmware while comprehensively elucidating the required technologies, specific implementation procedures, feasibility, and constraints, thus pioneering a new research domain in this area.
Note that, in this paper, the act of increasing the cost of vulnerability analysis is defined as ``obfuscation."
%By applying SmmPack, even on their own PC, attackers would face significant challenges in retrieving and analyzing SMM modules.
%A detailed discussion of the tradeoffs between the costs and benefits gained by implementing SmmPack is provided in Section~\ref{sec:cost-trade-offs}.

The key technical concept of SmmPack involves sealing the key used for encrypting the SMM modules in the TPM, preventing attackers from extracting the key even on their own terminals. Key retrieval is restricted by the Platform Configuration Register (PCR), making it impossible to obtain the key after the OS has booted. Additionally, during the boot phase, if such SMM module performing SMRAM dump is added, the PCR value changes, preventing key acquisition. As a result, attackers are limited to methods such as cold boot attacks involving memory transplantation and DMA within an extremely small time frame, forcing them to resort to costly methods\footnote{The magnitude of these costs is discussed in Section \ref{subsec:difficulties}.}. Furthermore, these methods may not be feasible depending on the presence of other security mechanisms.

% The key technical concept of SmmPack is illustrated below. In our proposed scenario, Original Equipment Manufacturers (OEMs) set up SmmPack before shipping computers to their users. 
% During this process, the SMM modules are encrypted using a key stored in the TPM. The key is unique to each BIOS implementation, meaning that BIOS with identical content can have the same key\footnote{The legitimacy of this implementation is discussed in Section \ref{sec:discussion}.}. By storing the encryption key in the TPM, even if an attacker obtains the SMM module from the SPI flash, they cannot decrypt or analyze its content unless they also obtain the key. Acquisition of the key is limited by the TPM's Platform Configuration Register (PCR). This restriction ensures that the key is only accessible during specific boot timings within the boot phase while preserving the integrity of the UEFI firmware.
% An important point to note here is that SmmPack assumes that the phase before Driver Execution Environment (DXE) is protected from being compromised by security mechanisms such as the Intel Boot Guard~\cite{bootguard}. This prevents key extraction by adjusting the PCR value in the early phases.

% 貢献
% In this study, we first defined the threat model for SmmPack, along with the corresponding requirements that must be fulfilled. We then implemented a prototype based on the requirements and evaluated its effectiveness \footnote{As an artifact of this paper, the implementation code of SmmPack is shown at \cite{github}.}.
In this study, we first defined the threat model for SmmPack. Then, we implemented a prototype and evaluated its effectiveness \footnote{As an artifact of this paper, the implementation code of SmmPack is shown at \cite{github}.}.
We conducted an assessment to determine whether SmmPack can effectively prevent various methods of acquiring SMM modules that attackers might attempt.
% Additionally, we demonstrate that the packed SMM modules function correctly without any loss of functionality and that the boot time and BIOS size overheads are both within practical limits.
Additionally, we demonstrate that the boot time and BIOS size overheads are both within practical limits.
Finally, we explain the procedures for managing and adopting SmmPack along with the process entailed in implementing BIOS updates and substantiate the feasibility of incorporating SmmPack.

The contributions of this study have been summarized as follows:
\begin{mybullet}
    \item We presented SmmPack, the first obfuscation framework developed for platform firmware, which is uniquely positioned as a pioneering solution that protects SMM modules from attacker vulnerability analysis.
    \item We demonstrated the effectiveness of SmmPack as a defense against various means of SMM module acquisition.
    \item We showed that the impact of SmmPack on the system's performance is minimal, indicating the practicality of its implementation.
    \item We clarified the management and deployment methods of SmmPack, as well as the procedure for applying BIOS updates, and demonstrated that the implementation of SmmPack is realistic.
\end{mybullet}
% }

\section{Background}
\label{sec:background}
In this section, we first provide an overview of UEFI, SMM, and TPM.
%We explain about UEFI since SMM module is part of the UEFI firmware.
Next, we explain what attackers can achieve by attacking SMM and discuss the types of vulnerabilities in SMM.
%To escalate privileges to SMM, it is necessary to identify vulnerabilities within SMM modules. Therefore, we describe CVEs in SMM modules.
Finally, we explain the security mechanisms that currently exist in UEFI.
%Finally, we organize the necessary conditions for conducting an attack on SMM.

\subsection{UEFI}
The Unified Extensible Firmware Interface (UEFI) is a set of specifications maintained by the Unified EFI Forum~\cite{uefi} that defines the interface between the platform firmware and the OS. A UEFI-compliant BIOS comprises numerous UEFI modules, most of which reside within a Serial Peripheral Interface (SPI) flash memory chip. The UEFI specifications partition the boot process into seven distinct phases, with the following five phases representing the most important stages of the procedure:

\begin{mybullet}
    \item SEC (Security): The initial code running and is the root of trust of the system.
    \item PEI (Pre-EFI Initialization): Initializes permanent memory and handles the different states of the system.
    \item DXE (Driver Execution Environment): Execute drivers that initialize platform components.
    \item BDS (Boot Device Selection): Select the boot device and run the boot loader.
    \item RT (Runtime): The phase when OS executes. UEFI environment except runtime services are discarded.
\end{mybullet}

Firmware that runs earlier in the boot phase has a more platform-dependent implementation. The phase until BDS is called the boot phase, and the phase after the OS starts is called the runtime phase. In this study, we specifically focused on the DXE phase, as this is where the SMM environment is also set up. Most UEFI modules running in the DXE phase are known as DXE modules. UEFI modules running after the PEI phase are usually in the PE format.

% UEFI has an important data structure called the UEFI system table, which allows access to various system configuration information and UEFI services. UEFI services are divided into two categories: boot services and runtime services. Boot services are only available during the boot phase, while runtime services are available during both the boot phase and the runtime phase. When the OS loader or OS kernel calls the ExitBootServices function of the boot service, the system switches to the runtime phase and boot services become unavailable.

% One of the services provided by the UEFI runtime services is the UEFI variable service, which allows storing various data in key-value format, such as the configuration of boot order or keys used in secure boot. The actual variable data is stored inside the SPI flash, with attributes that describe the visibility and persistence of each variable. If the variable has the RUNTIME\_ACCESS attribute, it can be read and written while the OS is running.

UEFI defines a standardized format for EFI firmware storage devices abstracted into Firmware Volumes (FVs). Typically, a BIOS contains multiple FVs, with one FV containing DXE modules, including SMM modules, and another FV containing PEI modules. UEFI modules in the DXE phase are individually dispatched by the dispatcher, which implies that no modules run in parallel. The dispatcher enumerates through the FVs to locate UEFI modules and execute them. FVs can include apriori files listing module GUIDs. Modules in this list run first, following the order of the file.
%The file system used is typically the Firmware File System (FFS). FFS is consisted of multiple EFI\_SECTION files, including PE32 image section which is the actual UEFI module executable.
% The types of EFI\_SECTION of DXE modules are:
% \begin{itemize}
%     \item DEPEX (DXE dependency section)
%     \item PE32 image section
%     \item User interface section
%     \item Version section
% \end{itemize}
% PE32 image section includes the actual UEFI module executable and DEPEX shows the GUIDs of protocol which should be installed before this DXE module.

EDK2 is a full implementation of the UEFI specification developed by the open-source Tianocore project \cite{edk2}. It is also a UEFI module development environment, which we used to build SmmPack.

\subsection{SMM}
The System Management Mode (SMM) is the highest-privileged operating mode of x86 and x86-64 processors, intended for use by BIOS firmware to handle low-level system management operations such as power management, system hardware control, or proprietary OEM-designed code. SMM can only be entered through a System Management Interrupt (SMI) and can only be exited with an RSM instruction or reboot. UEFI modules that run in SMM are called SMM modules, and these modules prepare SMI handlers that perform low-level system management operations.

\noindent{\bf SMRAM.}
SMM modules are loaded into a separate memory region called SMRAM, which can only be accessed during SMM and cannot be read via Direct Memory Access (DMA). SMM is typically initialized during the DXE phase. In this phase, SMM modules are also loaded and executed from their entry points. At the end of the DXE phase, the SMRAM is locked by setting Model Specific Registers (MSRs), which control the SMRAM. This locking makes these MSRs read-only and restricts the visibility of SMRAM solely to SMM.

\noindent{\bf SMM System Table and Protocols.}
SMM has a data structure called the SMM system table that allows access to the various functions used in most SMM modules. One of the functionalities provided by the SMM system table is the protocol. When a certain SMM module registers its provided function as a protocol in the system, other SMM modules can utilize that function. Protocols primarily consist of a GUID and a protocol interface structure. When the protocol is used, functions such as SmmLocateProtocol of the SMM system table first find the protocol interface structure based on the GUID. The specified function is then called, referencing the function pointers in the protocol interface structure. The protocol interface structure is usually defined in the data section of the SMM module that installed that protocol. The actual functions pointed to the protocol interface structure are also defined inside the SMM module.

\noindent{\bf CommBuffer.}
There are several ways to communicate data between non-SMM and SMM, the most common of which is to use the SMM communication buffer (CommBuffer) via the SMM communication protocol or the Advanced Configuration and Power Interface (ACPI) table. When using CommBuffer, SMI handlers should call the SmmIsBufferOutsideSmmValid function to validate the data. This is because if the data passed from the CommBuffer point to the data inside SMRAM, the non-SMM code can manipulate the data inside SMRAM.

\subsection{TPM}
Trusted Platform Module (TPM) is a secure cryptoprocessor designed to safeguard the artifacts used for platform authentication. TPM specification has been developed by the Trusted Computing Group (TCG)~\cite{tcg}. Currently, TPM 2.0 is the latest version of the specification. In this study, we focused primarily on TPM 2.0, with no discussion of TPM 1.2, its predecessor. TPM is equipped with numerous anti-tampering mechanisms that enable it to effectively resist a wide array of hardware and side-channel attacks.

\noindent{\bf PCR.}
One of the essential features of TPM is the Platform Configuration Register (PCR), which provides a method for measuring the state of software. The value stored in the PCR is a hash of the software code or configuration data and can only be updated through an extend (or reset) operation. By referencing the PCR value, the platform firmware ensures that no code executed during the boot phase is altered.
%The underlying principle is that any executable code is measured and extended into the PCR before it is executed.
If an attacker attempts to run a tampered UEFI module during the boot chain, the change in PCR values can detect tampering.
The extend operation hashes the concatenated value of the new measurement and the current PCR value.
% The extend operation is formulated as follows:
% \begin{equation}
%     PCR[i] = H(\,PCR[i]\:||\:m\,),
% \end{equation}
% where $PCR[i]$ denotes the $i$-th PCR value, $H()$ symbolizes the hash function, and $m$ represents the new measurement of the forthcoming program to be executed. 
% The $||$ operation signifies the concatenation of two byte arrays.
A standard TPM comprises 24 PCRs, each designated to store distinct types of measurements. In this study, we focused solely on PCR0, where the system firmware measurements are stored. The DXE phase measurement in the EDK2 framework was conducted at Firmware Volume (FV) granularity using the Tcg2Pei module during the Pre-EFI Initialization (PEI) phase~\cite{fvunitmeasure}.

\noindent{\bf Session.}
%TPM has two key concepts: authorizations and sessions. Authorizations pertain to access control information that can be associated with various entities within the TPM, whereas sessions serve as a means of authorization and maintain state between consecutive commands.
% Many TPM commands require the use of sessions that can be divided into the following three types:
% \begin{itemize}
%      \item Password: Include a password in the authentication data structure within the TPM command.
%      \item HMAC: Instead of exchanging a password each time, exchange the HMAC with a nonce to perform more secure authorization within the TPM.
%      \item Policy: Created on top of the HMAC session, more secure authorization is achieved by combining conditions that consider not only the password but also the TPM states and the states of external devices such as fingerprint readers.     
% \end{itemize}
Many TPM commands require the use of sessions that can be divided into the following three types: (1) Password, (2) HMAC, (3) Policy.
Among these three, policy session is the most powerful, allowing authentication based not only on passwords but also taking into account PCR values and other external information.
% Many TPM commands require the use of sessions that can be divided into the following three types: (1) Password sessions include a password in the authentication data structure within the TPM command. (2) HMAC sessions involve exchanging an HMAC with a nonce instead of a password, enabling more secure authorization within the TPM. (3) Policy sessions are created on top of the HMAC session, achieving more secure authorization by combining conditions that consider not only the password but also the TPM states and the states of external devices such as fingerprint readers.
In this study, we adopt a policy session.
%A policy session enables the integration of PCR values as an authorization factor.

\noindent{\bf Sealing.}
Storing secrets in TPM with the policy session using PCR is called sealing.
Sealing can confine the accessibility of stored data to specific instances during the boot phase when platform integrity is assured.

\noindent{\bf NV space.}
A certain amount of nonvolatile (NV) space is available in the TPM for user-configured storage. Authorization can be applied to the data stored in NVRAM, allowing the data to be sealed inside. The data placed in NVRAM can be accessed through a handle called the NV index.

% \noindent{\bf Control Domain.}
% Three entities can control the TPM: the platform firmware, the platform owner, and the privacy administrator. In most cases, the platform owner is the same as the privacy administrator.
% %Each entity has a different authorization value, authorization policy, and seed to control TPM.
% The OS corresponds to the owner, and UEFI BIOS corresponds to the platform firmware, resulting in distinct control domains. TPM data, which belongs to UEFI, does not get affected by the OS. For example, the NV Index, which belongs to UEFI, does not get cleared when the OS initializes the TPM using the TPM2\_Clear command.

\subsection{Vulnerabilities and attacks in SMM}
By escalating privileges to SMM, attackers can arbitrary write to SPI flash, bypass hypervisor-based security mechanisms~\cite{vbsbypass,vbsbypass2}, and provide stealthy functions at OS runtime~\cite{deitybounce}. Moreover, all attacks that can be done by infecting other UEFI modules are also possible from SMM modules.

Various types of vulnerabilities that can trigger the above attacks have been identified in SMM modules, including SMM callout \cite{smmcallout,smmcallout2}, confused deputy \cite{confuseddeputy2018}, and buffer overflow \cite{smmbof,smmheapbof}. 
Most CVEs are caused by implementation errors in SMM modules. However, attacks against SMM are not widely known and obfuscation to hide these vulnerabilities is not currently performed.

\subsection{Existing security mechanisms}
Existing security mechanisms for preventing attacks on UEFI include secure boots and an Intel Boot Guard.

\noindent{\bf Secure boot or Verified boot.}
These security mechanisms are designed to ensure the system integrity. The central idea is to ensure that the system firmware has not been tampered with by verifying the hash of the firmware. Therefore, it is impossible to prevent attacks that do not tamper with the code. 
%For example, writing a malicious payload to CommBuffer does not compromise code integrity; therefore, a secure boot cannot prevent such attacks.
Secure boot mainly checks the integrity of the firmware after PEI and operates with trust in the OEM platform firmware before DXE.
%In addition, a secure boot can be toggled on or off by the user, usually from the BIOS setup screens.

% \noindent{\bf SPI flash write protection.}
% Various methods exist to write-protect SPI flash.
% Firstly, there are five Protected Range Registers (PRRs) which can specify read and write permissions for one memory address space per register. This always affects when using the MMIO registers of the SPI flash, and it is not possible to bypass PRRs even with SMM privileges. The FLOCKDN register can be used to lock the reading and writing of the PRR register itself, and FLOCKDN cannot be disabled without a reboot. Since PRR and FLOCKDN are powerful but inflexible,
% One such method is using the BIOS\_CNTL registers. BIOS\_CNTL.BIOSWE can enable or disable SPI flash write operations, and reading and writing to this register can be locked by BIOS\_CNTL.BLE. This "lock" mechanism works such that when 1 is written to BIOSWE, BIOSWE becomes 1 once, but then an SMI occurs, and it is returned to 0 by the SMI handler. Therefore, an attacker who can run the code in SMM can disable this.

\noindent{\bf Intel Boot Guard.}
%While the secure boot verifies the integrity of the firmware after the PEI phase, the
Intel Boot Guard \cite{bootguard} verifies the integrity of both the SEC and PEI phases. The Intel Boot Guard is a hardware-based technology that cannot be disabled by users. In AMD, similar functionality is implemented as a Platform Secure Boot (PSB)~\cite{amdpsb}, which also cannot be disabled by the user.

% \subsection{Conditions for attacking SMM}
% In order to execute code in SMM privilege using vulnerabilities such as mentioned in the above CVEs, attackers must first obtain and analyze the SMM modules to find the vulnerability. Implementation of SMM modules varies depending on the BIOS version and PC model, making it necessary for attackers to analyze the BIOS of the target PC rather than their own. Attackers can obtain the same BIOS by downloading the BIOS update file or purchasing the same PC if they can determine the version of the BIOS or the model of the PC. Most of the existing security mechanisms are designed to ensure system integrity and are not equipped to effectively counter attacks that exploit vulnerabilities in SMM modules. As a result, if an attacker discovers vulnerabilities in the SMM modules of a target BIOS, they can easily exploit these vulnerabilities to carry out attacks.
\section{Threat Model}
\label{sec:threatmodel}

The purpose of SmmPack is to increase the cost of vulnerability analysis of SMM modules by attackers. Therefore, the threat model of SmmPack is defined as ``attackers aiming to discover vulnerabilities in SMM modules."

SmmPack is not intended to prevent attacks towards SMM. Its purpose is to increase the cost of vulnerability analysis conducted prior to those attacks. To the best of our knowledge, no research has been conducted on the encryption or obfuscation of system firmware, including SMM modules. Consequently, there are no existing threat models focused on the vulnerability analysis of SMM modules. 
Therefore, there is a need to define a threat model targeting attackers aiming to conduct vulnerability analysis on SMM modules.

%The threat model targeting the vulnerability analysis of SMM modules differs from the threat model attacking SMM. Attackers with the intention of attacking SMM must perform their attacks on the victim's PC, not on their own PC. However, 
Attackers who intend to analyze SMM modules to identify vulnerabilities can obtain the modules from their own PC. Therefore, the threat model must consider the Man-at-the-End (MATE) model~\cite{manattheend}. In general, research considering this model finds it challenging to completely prevent attackers from conducting vulnerability analysis. Instead, the focus is on increasing the costs for attackers.

It is necessary not only to address attacks on SMM but also to take measures for the vulnerability analysis of SMM modules.
This is because many attacks that exploit vulnerabilities in the SMM modules cannot be prevented using existing security mechanisms. Attacks such as buffer overflow through CommBuffer~\cite{smmbof,smmheapbof} and SMM callout attacks~\cite{smmcallout,smmcallout2} do not require tampering with the SMM module. Thus, such security mechanisms like secure boot is ineffective.
%The most reliable approach to thwarting such attacks is to implement SMM modules without vulnerabilities. However, 
It is highly challenging for developers to comprehend all SMM-related attacks and implement all modules without any vulnerabilities.
Furthermore, even if vulnerabilities are identified, BIOS has the characteristic of having a significantly longer period until a patched version is released compared to regular software. Therefore, increasing the cost of vulnerability analysis is an important countermeasure.

% 脅威モデルの提案
In this study, the threat model was classified into the following three classes based on the attacker's accessible scope.
In this threat model, attackers aim to obtain decrypted SMM modules and the key. Vulnerability analysis is also prevented by preventing the acquisition methods.
Class 3 attackers have greater access than class 1 attackers; however, they must employ relatively expensive methods to obtain SMM modules.
%Furthermore, the assumptions to be considered when evaluating SmmPack with this threat model are explained in Section \ref{sec:requirements}.

~\\
\noindent{\bf Class 1 attacker}: attacker with access to only BIOS update file.

\noindent{\bf Class 2 attacker}: attacker with arbitrary software code execution above OS.

\noindent{\bf Class 3 attacker}: attacker with hardware access and equipment.
\\

\noindent{\bf Class 1 attacker.}
The lowest cost way to obtain SMM modules is to acquire them from BIOS update files. These BIOS update files can be freely downloaded from various manufacturer websites, making it cheaper than purchasing a device.

\noindent{\bf Class 2 attacker.}
A Class 2 attacker aims to obtain SMM modules by purchasing a PC containing the BIOS image they want to analyze and obtaining SMM modules from the software running on the OS. 
This attacker can attempt to dump SPI flash from the software, attempting to access keys through software-based access to the TPM using tools such as tpm2-tools~\cite{tpm2-tools}, and trying to obtain decrypted SMM modules loaded in memory through memory dumps.

\noindent{\bf Class 3 attacker.}
A Class 3 attacker can attempt to acquire SMM modules using various methods, ranging from low-cost approaches, such as dumping the BIOS image from the SPI flash chip, to more costly methods, such as acquiring the decrypted SMM modules from memory via cold boot attacks, which require considering various hardware-specific security mechanisms. Attackers can also execute arbitrary SMM modules by flashing the SPI flash.

% Compared to the class 2 attacker, the class 3 attacker has less flexibility due to the lack of abstracted services offered by the OS and debugging capabilities. The class 3 attacker also needs some special hardware devices to flash the BIOS, making their approach more costly. 
% The class 3 attacker employs more hardware-based methods such as cold boot attacks to obtain SMM modules. In contrast to the class 3 attacker, the class 3 attacker needs to deal with hardware-specific mechanisms like memory scrambling, making their approach even more costly.

% Currently, there is no security mechanism for preventing the acquisition and analysis of UEFI modules. Hence, attackers ranging from Class 1 to Class 3 can obtain SMM modules by using any of the methods mentioned above.

%By evaluating the extent to which attackers of ranging from class 1 to class 3 can be mitigated, it is possible to assess the increase in cost concerning vulnerability analysis of attackers.

When evaluating using this threat model, the following assumptions are made: (1) the TPM module is tamper-resistant, (2) attacks on cryptographic algorithms themselves are considered out-of-scope, and (3) the integrity of the SEC and PEI phases is hardware-protected by the Intel Boot Guard or AMD PSB.
Therefore, in this paper, we consider vulnerabilities in TPM itself, weaknesses in cryptographic algorithms, and tampering of SEC and PEI phases to be out of scope.
\section{Implementation} \label{sec:implementation}
In this section, we explain the implementation of SmmPack. First, we provide an overview of SmmPack, followed by a description of the preparation of the protocol used for the decryption, sealing, and unsealing of keys, and details of the packer. In Section \ref{sec:discussion}, we discuss the validity of the design.

\subsection{SmmPack overview}
\begin{figure}[tb]
  \centering
  \includegraphics[width=0.75\linewidth]{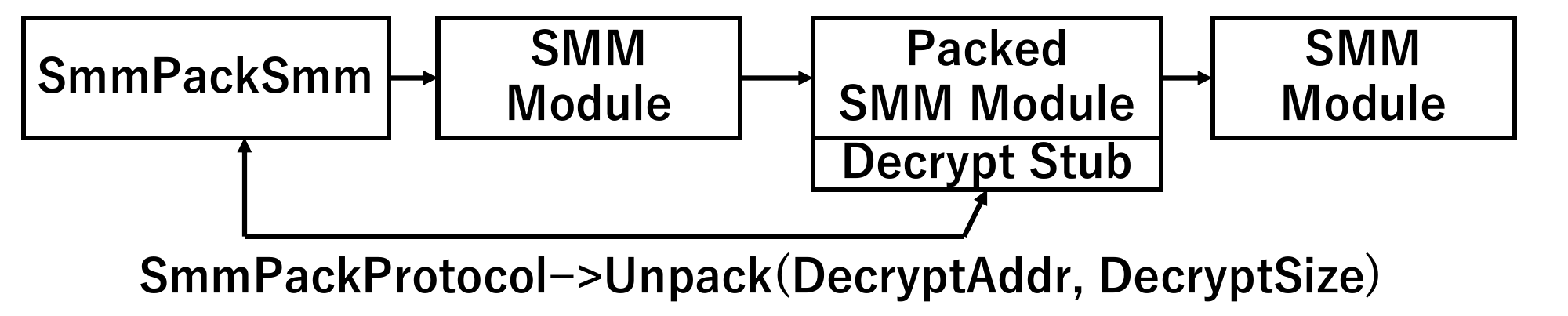}
  \caption{SmmPack overview.}
  \label{fig:smmpack-overview}
\end{figure}

Figure \ref{fig:smmpack-overview} illustrates an overview of SmmPack. SmmPack packs the SMM module by encrypting its code section and adding the decrypt stub at its end. When the packed SMM module is executed, the execution starts from the decrypt stub, which calls the Unpack function of SmmPackProtocol that the SmmPackSmm module installs. Subsequently, SmmPackSmm will decrypt the code section of the packed SMM module using a key stored inside TPM.

\begin{figure}[tb]
  \centering
  \includegraphics[width=0.9\linewidth]{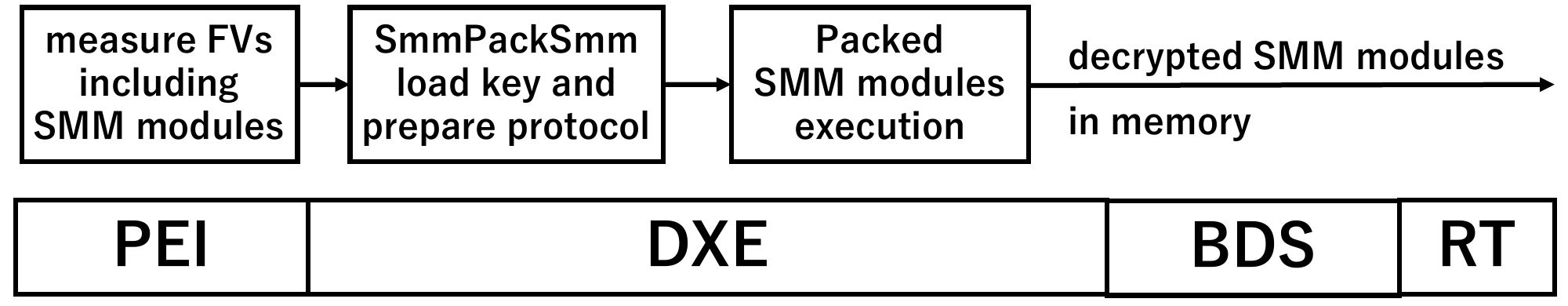}
  \caption{High level core operations in SmmPack.}
  \label{fig:highlevel}
\end{figure}

Figure \ref{fig:highlevel} shows the key operations related to SmmPack at a high level. SmmPackSmm loads the decryption key from the TPM in the early DXE phase. Because the key is sealed, the PCR value must be a specific value. The PCR value was calculated in FV units in the PEI phase. After SmmPackSmm installs the SmmPackProtocol, the packed SMM modules are executed, and once executed, they remain decrypted in SMRAM throughout the runtime.

Table \ref{tab:component} lists the components used in SmmPack, and Table \ref{tab:env} presents the environmental information. It is assumed that the packing of SMM modules will be performed by the OEM before shipment by executing the packer program only once. The keys are unique values for each BIOS implementation and are sealed before shipment to the non-volatile area of the TPM.

\begin{table}[tb]
\caption{SmmPack components.}
\label{tab:component}
\centering
\begin{tabular}{p{23mm}|l}
\hline
name & description \\
\hline
SealKeyDxe & Seal key into the TPM \\
SmmPackSmm & Provides protocols for unpacking \\ 
smm-packer & Packer program \\
tiny-AES-c \cite{tinyaes} & Library for AES \\
\hline
\end{tabular}
\end{table}

\begin{table}[tb]
\caption{Experiment setup.}
\label{tab:env}
\centering
\begin{tabular}{l|l}
\hline
name & description \\
\hline
Visual Studio 2019 & for building packer program/{\tt C++} \\
EDK2 & for building UEFI modules \\ 
UP Squared Pro Atom 04/64 \cite{up2pro} & for executing UEFI modules \\
CPU & Intel Atom® x7-E3950 \\
BIOS & UNAPAM22 \\
TPM & Infineon SLB9665 \\
\hline
\end{tabular}
\end{table}

\subsection{SmmPackSmm}
SmmPackSmm mainly performs the following two processes: (1) Unsealing the key from TPM and (2) Registering a protocol for decryption.
SmmPackSmm is an SMM module that installs an SMM-specific protocol called SmmPackProtocol for unpacking packed SMM modules. When the packed SMM module is dispatched, it starts execution from the decrypt stub added by our packer, and the decrypt stub calls the Unpack function of SmmPackProtocol to decrypt itself. To pack more SMM modules, the protocol must be installed before any packed SMM module is executed. For this purpose, the GUID of SmmPackSmm should be written to the apriori file of its FV. Despite multiple FVs, the protocol can be used across FVs. Therefore, if SmmPackSmm is registered in the apriori file of the initially executed FV, the modules in other FVs can also be packed.

\noindent{\bf Unsealing key from TPM.}
\begin{figure}[tb]
  \centering
  \includegraphics[width=0.75\linewidth]{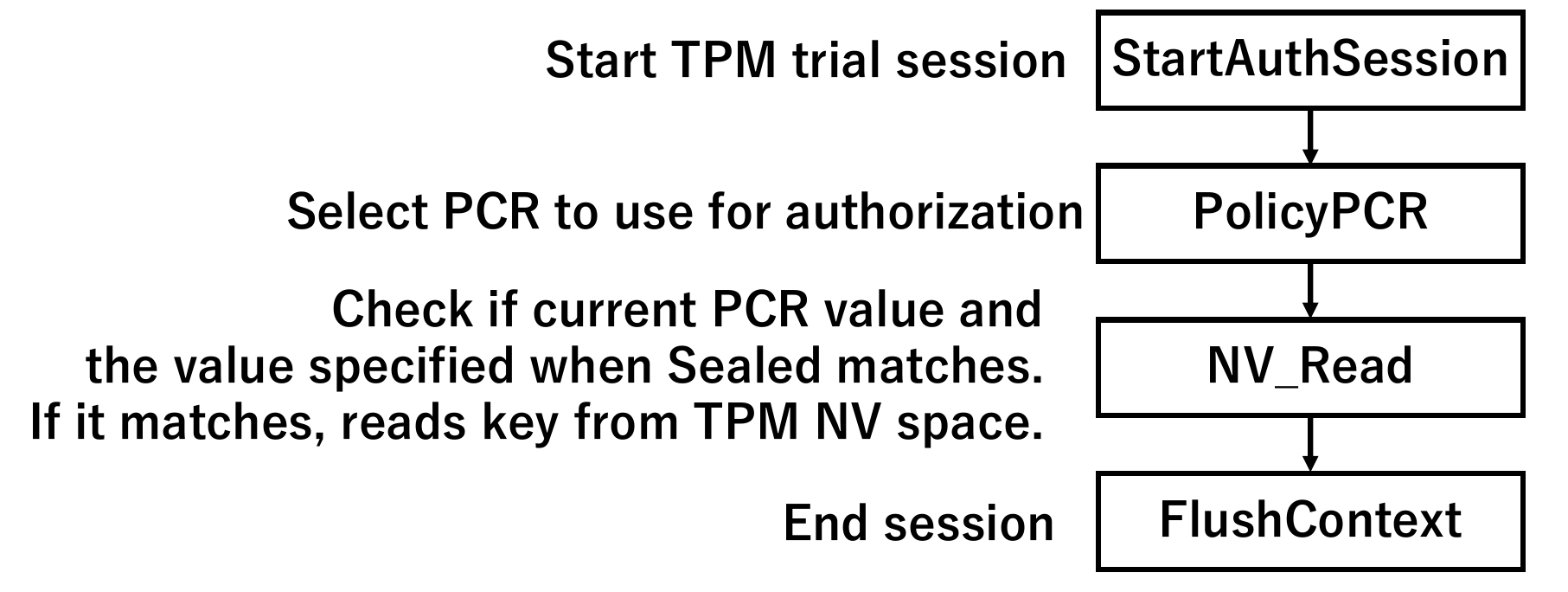}
  \caption{Key unsealing commands.}
  \label{fig:unseal}
\end{figure}
To unseal the decryption key, SmmPackSmm sends commands to the TPM as shown in Figure \ref{fig:unseal}, using the Tpm2DeviceLib library in EDK2. Tcg2Protocol and similar protocols cannot be used with SmmPack because they are not available until the DXE module that installs them is executed. First, TPM2\_StartAuthSession is used to start a policy session, followed by TPM2\_PolicyPCR to select the PCR to use for authorization for this session. %In this case, PCR0 was selected from the 24 available PCRs.
Finally, TPM2\_NV\_Read is used to read the key from the nonvolatile storage of the TPM, with verification of the current PCR value to ensure that it matches the enrolled PCR value specified when the key was sealed. The key is then saved as a global variable in SmmPackSmm. It is also possible to perform decryption inside the TPM without loading the key into memory. However, this method requires SmmPackSmm to send the encrypted data (the entire text section of the packed SMM modules) to the TPM, which significantly increases the execution time. Loading the key into memory does not increase the attack vectors in the context of this study, if the key could be read from memory, so too can the unpacked SMM modules.

\noindent{\bf Registering a protocol for decryption.}
After the key is unsealed, SmmPackSmm installs a protocol that decrypts the packed SMM modules. The protocol is a table of function pointers, and in SmmPack, only one function, Unpack, is required. The Unpack function takes the base address and the size of the encrypted text section of SMM modules and decrypts it using the key stored in the global variable. The encryption algorithm used in SmmPack is AES-128 in CBC mode, which is implemented by using the open-source library tiny-AES-c. However, this can be any similar-strength symmetric algorithm or implemented using any other library. After the protocol is installed, the subsequent SMM modules can locate it using its GUID and call the Unpack function.

\subsection{Key sealing}
\begin{figure}[tb]
  \centering
  \includegraphics[width=0.72\linewidth]{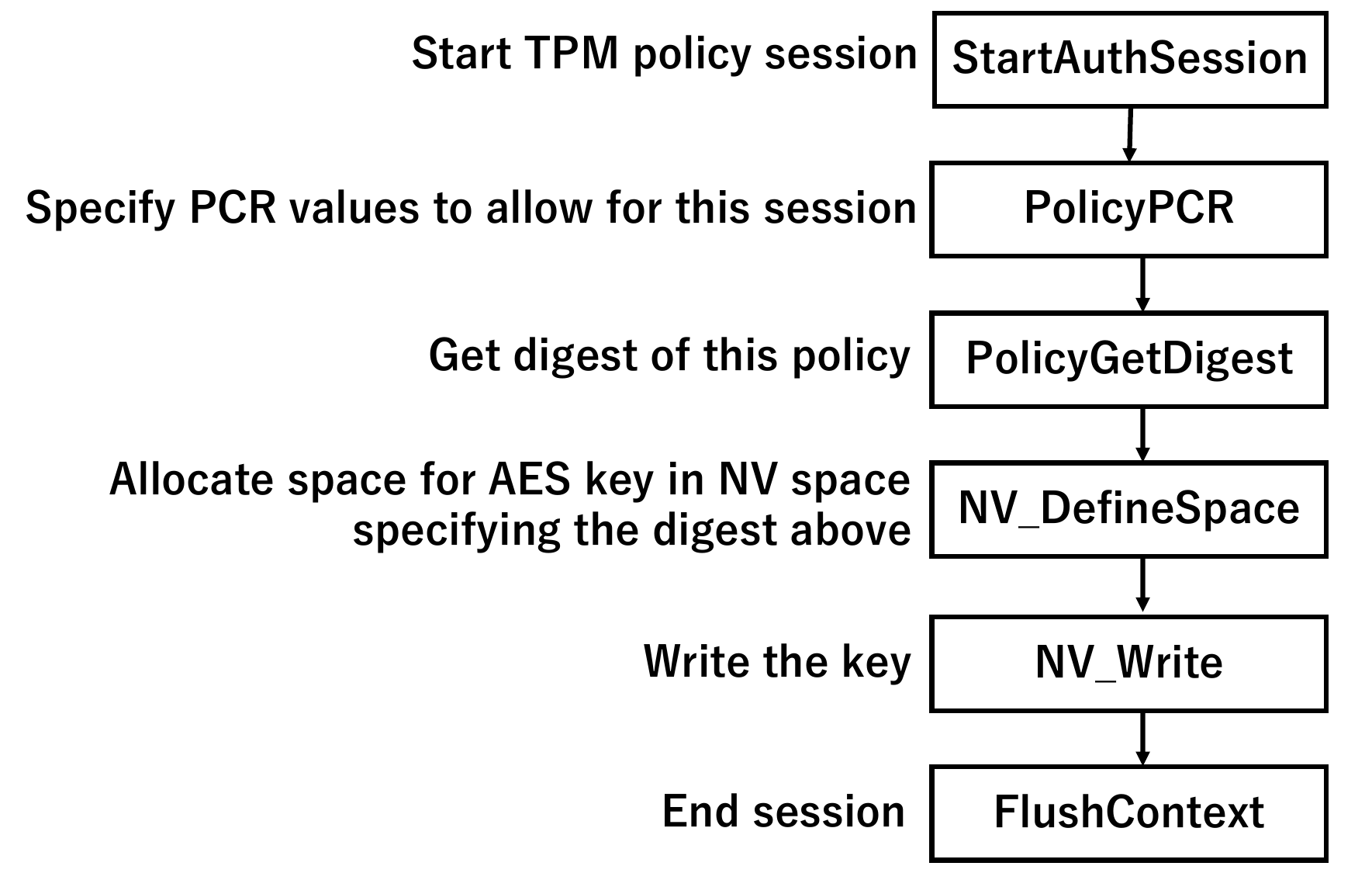}
  \caption{Key sealing commands.}
  \label{fig:seal}
\end{figure}

The key used by SmmPack must have a unique value for each BIOS implementation. SmmPack aims to prevent the analysis of SMM modules; therefore, if the BIOS code is the same, encryption using a different key is not required. Figure \ref{fig:seal} presents an overview of sealing a key to a TPM. TPM2\_NV\_DefineSpace allocates space to place a key inside the nonvolatile storage of the TPM. This command is sent with a digest of the policy obtained from TPM2\_PolicyGetDigest, which contains the information on the PCR value that can access this space. This information was specified by TPM2\_PolicyPCR on a trial session initiated by TPM2\_StartAuthSession. SealKeyDxe, which performs this operation, should be executed before packing any SMM module with SmmPack. Alternatively, this process can be performed using other tools, such as tpm2-tools \cite{tpm2-tools}.

\subsection{Packer design}
\begin{figure}[tb]
  \centering
  \includegraphics[width=0.72\linewidth]{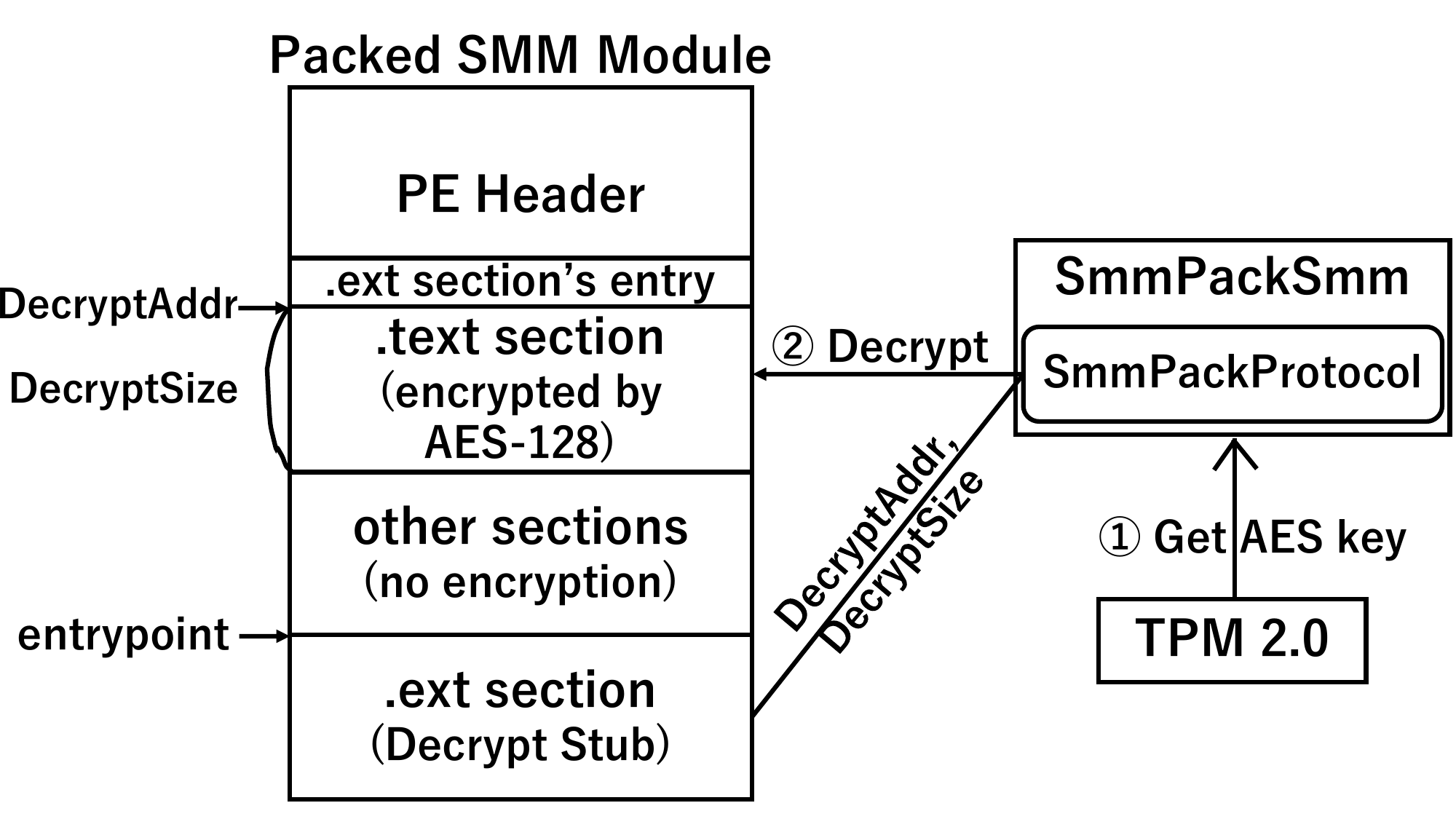}
  \caption{Structure of packed SMM module.}
  \label{fig:packer}
\end{figure}
The packer program was implemented separately from the other components as an application running on top of the OS. This is also implemented using tiny-AES-c because the algorithms used for encryption and decryption must be the same. Figure \ref{fig:packer} shows the structure of the packed SMM module. The packer first encrypts the text section of the SMM module with AES. Then, the packer adds the ext section (decrypt stub) containing the decryption code at the end of the SMM module. The entry point of the module should also be moved to the decrypt stub.
SmmPack encrypts only the text section because the purpose of SmmPack is to prevent vulnerabilities from being discovered by analyzing the codes of the SMM module.

% An additional entry must be added to the PE header to add a section to a PE file. This requires pushing back all other PE sections and adjusting all the related header values. If the SMM module being packed installs one or more protocols, it is necessary to shift the addresses of the functions in those protocol interface structures\footnote{We note that this process can be automated, and for a sample implementation, readers can refer to our source code \cite{github}.}.
%Also, if the protocol interface structure is included in the text section to be encrypted, it has been confirmed that execution of the packed SMM module will stop. Therefore, the SMM modules that are packed need to have the protocol interface structure in a section other than the text section.
%\section{Security analysis (Req 1)} \label{sec:analysis}
\section{Security analysis} \label{sec:analysis}
% In this section, we present the evaluation results of the first requirement defined in Section \ref{sec:requirements}. Utilizing the threat model defined in Section \ref{sec:threatmodel}, we performed an evaluation and presented the overall results in Table \ref{tab:alleval}.
In this section, we assess the extent to which SmmPack increases the cost of vulnerability analysis, utilizing the threat model defined in Section \ref{sec:threatmodel}. We presented the overall results in Table \ref{tab:alleval}.

In summary, SmmPack can completely prevent all acquisition methods for Class 1 and Class 2 attackers and can restrict Class 3 attackers' acquisition methods to only cold boot attacks and DMA.
However, neither of these acquisition methods may be applicable because of the existence of specific security mechanisms. Even if these security mechanisms were not present, performing these acquisition methods against SmmPack would still be a highly cost-intensive task. Detailed explanations are provided in Subsection \ref{subsec:difficulties}. 
Hence, using SmmPack, the avenues for attackers to acquire SMM modules are narrowed down to methods with a significantly higher cost.

% In summary, SmmPack can defend against all acquisition methods except for obtaining SMM modules through a cold boot attack by a class 3 attacker. However, as discussed later, cold boot attacks have become highly cost-intensive methods in recent years. Thus, the use of SmmPack demonstrates its ability to significantly increase the cost of vulnerability analysis for attackers.
% Regarding the case of obtaining through a BIOS update file, this will be explained in Section \ref{sec:discussion}.

\begin{table}[]
\caption{Threat models and attacks SmmPack can address (O represents acquisition methods that can be prevented by SmmPack, while X represents methods that cannot be prevented by SmmPack.).}

\label{tab:alleval}
\centering
\begin{tabular}{lll}
\hline
class 1                 & BIOS update file                    & O \\
\hline
class 2                 & SPI flash dump (software)           & O \\
                        & key read from TPM (software)        & O \\
                        & memory dump from OS                 & O \\
\hline
class 3                 & SPI flash dump (hardware)           & O \\
                        & key read from TPM (hardware)        & O \\
                        & SMRAM read from SMM module          & O \\
                        & DMA                                 & X \\
                        & coldboot attack (same device)       & O \\
                        & coldboot attack (memory transplant) & X \\
\hline
\end{tabular}
\end{table}

\subsection{Class 1 attacker}
\noindent{\bf BIOS update file.}
Although changes to the BIOS update method are necessary, distributing a BIOS image file containing SMM modules in an encrypted state would prevent attackers from analyzing them because they would not possess the decryption key.
%Therefore, it can be concluded that SmmPack is effective against Class 1 attackers.
Further details regarding the BIOS update method upon adoption of SmmPack are discussed in Section \ref{sec:discussion}.

\subsection{Class 2 attacker}
\noindent{\bf SPI flash dump (software).}
In this case, the most reliable method for obtaining SMM modules is to dump the BIOS from the SPI flash, which can be performed using tools such as CHIPSEC \cite{chipsec} or RWEverything \cite{rweverything}. With SmmPack, the obtained modules are encrypted; thus, attackers cannot analyze them.

\noindent{\bf key read from TPM (software).}
Even if the modules are encrypted, an attacker capable of retrieving the key can still analyze their content. It is possible to access the TPM from the software running on top of the OS using tools such as tpm2-tools \cite{tpm2-tools}. However, it is not possible to unseal the key from the TPM on top of the OS because the PCR value has changed by then\footnote{To ensure certainty, it is also possible to extend PCR0 in SmmPackSmm after unsealing the key.}. Therefore, even if a packed SMM module is obtained on top of the OS, it cannot be analyzed.

\noindent{\bf memory dump from OS.}
Because the packed SMM modules are placed in memory in a decrypted state after execution, attackers can obtain the decrypted modules through a memory dump without the need to acquire both the encrypted SMM module and the key. Although SMM modules exist in memory during runtime, they are located in SMRAM, making it impossible for the software on the OS to read. SMRAM is accessible only by SMM and SMM is entered only by SMI. Consequently, SMRAM can only be dumped from SMM modules. Hence, Class 2 attackers cannot obtain SMM modules using memory dumps.

\subsection{Class 3 attacker}
\noindent{\bf SPI flash dump (hardware).}
%In this case, the most reliable method for obtaining the SMM module is to directly extract it from the SPI flash chip.
Using SmmPack, SMM modules within the SPI flash are always stored in an encrypted state at any point in time. Therefore, this method by itself does not allow an attacker to analyze the SMM module.

\noindent{\bf key read from TPM (hardware).}
To decrypt the encrypted module, the attacker must obtain the key from the TPM. Note that the vulnerability of TPM's tamper-resistance feature is considered beyond the scope of this paper. Therefore, we assume that the physical retrieval of the key is prevented.

\noindent{\bf SMRAM read from SMM module.}
While attackers can disable secure boot on their PC and write an SMM module that performs an SMRAM dump, introducing any SMM modules will alter the PCR value. As a result, neither the key nor the decrypted SMM modules are loaded in the SMRAM by SmmPackSmm. PCR values are usually measured in the PEI phase on a per-FV basis before transitioning to the DXE phase~\cite{fvunitmeasure}. Therefore, even when running SMM modules in the early DXE phase, the PCR value changes.
%if an SMM module that reads the SMRAM is included in the FV, the PCR value would differ, and unsealing would fail.

\noindent{\bf DMA.}
An attacker can also read data from the memory via Direct Memory Access (DMA) without requiring any additional modules to write.
SMRAM is an area that cannot be accessed via DMA, so generally, this is not feasible.
However, the exact point at which SMRAM becomes inaccessible is after the SMRAM is locked. Consequently, performing DMA during the DXE phase after the execution of SmmPackSmm is possible until SMRAM is locked. This allows for the retrieval of decrypted SMM modules or keys.

% However, SMM modules cannot be obtained, as they are located in SMRAM which cannot be accessed even with DMA. One method to read SMRAM data through DMA is the "tich attack". The base address and range of SMRAM are specified using MSRs. By altering these MSRs, the range of SMRAM can be shifted. This enables conducting DMA on an area specified outside the original SMRAM range, allowing access to desired data stored in SMRAM. Normally, these MSRs are locked and set to read-only. However, SMRAM setup occurs early in the DXE phase and lockdown happens at the end of DXE. Therefore, it is possible to alter the MSRs during the DXE phase, potentially enabling tich attacks. Yet, modifying the MSRs requires writing DXE or SMM modules, leading to PCR value changes, rendering tich attacks unfeasible. As a result, it's not possible to use DMA to read decrypted SMM modules.
% DMA is a method that can be used if the PC has a DMA-capable port and the attacker can disable Intel VT-d (or IOMMU) \cite{intelvtd} on that platform.

Another possible method is a cold boot attack \cite{coldbootattack}, which takes advantage of the delay until the DRAM content is completely erased and the residual data are read out. Cold boot attacks can be further categorized into two methods based on their approach. The first method involves performing a cold reset within the same device and then reading the residual data. The second method involves freezing memory modules, transplanting them to another device, and then reading the residual data on that device.
%In either method, instead of reading the data of SMM modules placed in RAM during the current boot, attackers only need to read the residual data in RAM. Hence, this approach is not subject to PCR restrictions.

\noindent{\bf cold boot attack (same device).}
In this case, it is impossible to obtain the decrypted SMM module for two reasons.
% To carry out this attack, an attacker needs to follow steps similar to the following:
% \begin{enumerate}
%     \item Power down the PC after all SMM modules have been loaded (not using the OS shutdown function, but cutting off power by unplugging).
%     \item Freeze the memory modules using cold spray.
%     \item Write a DXE module that reads the residual data and store it in the SPI flash (configure it to run during the initial stages of the DXE phase, e.g., using apriori files).
%     \item Boot up the PC again.
% \end{enumerate}
To carry out this attack, an attacker must follow steps similar to the following: (1) Power down the PC after all SMM modules have been loaded, (2) Freeze the memory modules using cold spray, (3) Write a DXE module that reads the residual data and stores it in the SPI flash (configure it to run during the initial stages of the DXE phase, e.g., using apriori files), and (4) Boot up the PC again.
An essential consideration is that the module responsible for reading the residual data must run before the BDS phase. This is because the residual SMRAM data are overwritten with new data during the DXE phase. As a premise of SmmPack, as stated in Section \ref{sec:threatmodel}, SEC and PEI phases are protected at the hardware level to prevent tampering. Thus, attackers must read the residual data through a DXE module. Furthermore, if the DXE module is added, the PCR value changes, halting the boot process when SmmPackSmm is executed. Therefore, the DXE module must be executed during the initial stages of the DXE phase before SmmPackSmm is executed.

To insert a DXE module that reads the residual data, attackers must flash the BIOS of the system. Typically, flashing a BIOS requires several minutes to complete.
However, modern DDR3 and DDR4 memory modules struggle to retain data for more than a minute, even when adequately cooled~\cite{ddr3scramble,ddr4scramble}.
%Existing research reports that the data in DDR3 memory modules fully decays within 10 s, even with sufficient cooling~\cite{ddr3scramble}. Although precise data for DDR4 memory modules are lacking, they are generally described as having faster data decay than DDR3 modules~\cite{ddr4scramble}. 
Therefore, it is considered impractical to obtain residual SMRAM data through a cold boot attack on the same system.

Furthermore, most systems employ the Memory Overwrite Request (MOR) mechanism ~\cite{mor}, which zero-clears the memory when a clean shutdown is not performed, as in the case of a cold reset.
%This mechanism results in the memory being zero-cleared during the Power-On Self Test (POST) stage, especially when a clean shutdown by the OS is not performed, as in the case of a cold reset.
% TODO: MORが結構普及している事を示す文献を引用
For these two reasons, a cold boot attack on the same system cannot retrieve the decrypted SMM modules.

\noindent{\bf cold boot attack (memory transplant).}
In this scenario, an attacker can freeze the memory, transplant it to another PC containing the DXE module that reads the residual SMRAM data, and then boot.
%By doing this, the attacker can retrieve decrypted SMM modules from the residual data.
MOR bit is not effective, because performing a clean shutdown and setting the bit to zero on the target PC beforehand would prevent memory zero-clearing during POST.
%Performing a clean shutdown and setting the MOR bit to zero on the target PC beforehand would prevent memory zero-clearing during POST. As a result, in this case, the MOR mechanism would not serve as an effective countermeasure.

% Memory Overwrite Request mechanism \cite{mor} performed in the earliest Power-On Self-Test (POST) during the boot phase, or hardware-based full memory encryption \cite{memoryencryption} can be considered.

% In summary, the DXE driver could be obtained and analyzed via memory read in the BDS phase or DMA. However, if the device does not belong to the attacker, the acquisition can be prevented by the presence of a BIOS password, Intel VT-d, and the absence of a DMA-capable port. On the other hand, the SMM module cannot be decrypted by any means and satisfies the requirements perfectly.

\subsection{High cost of DMA and cold boot attacks} \label{subsec:difficulties}
Although SmmPack may not prevent DMA and cold boot attacks involving memory transplant, these methods may not be viable depending on the presence of specific security mechanisms. Furthermore, even without those, attempting these acquisition methods against SmmPack remains a highly challenging task.

\noindent{\bf DMA difficulties.}
There are two non-negligible reasons why the retrieval of decrypted SMM modules or keys using DMA is an extremely challenging task. First, the window of time between the initialization and the locking of SMRAM is very brief. Consequently, it is difficult to time the DMA operation. Attempts to insert a deadlock loop would alter the PCR values, preventing the loading of keys and halting the execution in SmmPackSmm. This limits the duration for which DMA can be performed, thereby increasing the cost for attackers.

Second, modern CPUs include various security mechanisms for DMA protection, including the DMA Protected Range (DPR)~\cite{dpr}. The memory region specified by DPR is inaccessible to DMA, and typically, DPR covers the memory areas used by SMRAM. If this security mechanism is active, attackers cannot use DMA to retrieve decrypted SMM modules or keys. Even if an attacker attempts to modify CPU registers to alter DPR settings, executing firmware to achieve this change requires changing the BIOS firmware, which alters PCR values. Consequently, if this security mechanism is enabled, DMA is unavailable.

\noindent{\bf Cold boot attack difficulties.}
Performing cold boot attacks requires consideration of various security mechanisms present in memory modules and system-on-chips (SoCs). DDR3 and later DRAM modules have a memory-scrambling feature that allows only scrambled data in memory, which is descrambled only within the SoC. 
%Although the detailed implementation of scrambling has not been publicly disclosed, existing research~\cite{ddr3scramble,ddr4scramble} reveals that a basic implementation involves XOR operations between a pseudo-random bitstream generated within the SoC using seed and memory addresses and plaintext data. This seed changes with every cold reset.
% If this is enabled and there is no option for users to disable it, attackers need to perform descrambling to read the residual data. 
Although descrambling of DDR3 and DDR4 DRAM modules are possible~\cite{ddr3scramble,ddr4scramble}, this will significantly increase the attacker's cost.
% Descrambling has been demonstrated in previous studies for both DDR3~\cite{ddr3scramble} and DDR4~\cite{ddr4scramble}. This suggests that cold boot attacks cannot be prevented even with scrambling enabled, although it can significantly increase the attacker's cost.

Moreover, certain SoCs employ technologies such as Intel's Total Memory Encryption (TME)~\cite{inteltme}, Total Memory Encryption Multi Key (TME-MK)~\cite{tmemk}, and AMD's Secure Memory Encryption (SME)~\cite{amdsme}. With these features enabled, data are always stored in an encrypted state in the memory and decrypted within the SoC only.
%Although memory scrambling was originally devised for power efficiency and noise reduction purposes and not primarily for security, these technologies were developed as security mechanisms to counter cold boot attacks.
Cold boot attacks are thwarted when these features are active. Although these mechanisms can often be disabled by the user from the BIOS setup screen, if these features are designed such that they cannot be disabled by design, attackers will be unable to perform cold boot attacks.

%\input{07_func}
%\section{Performance evaluation (Req 3)}
\section{Performance evaluation}
% In this section, we evaluate the performance of SmmPack and demonstrate that SmmPack satisfies the third requirement defined in Section \ref{sec:requirements}. We measured the boot delay in clock cycles when SmmPack was introduced and compared it with a normal boot. Also, we measured the extent to which the size of the entire BIOS increases by introducing SmmPack.
In this section, we evaluate the impact of SmmPack on system performance and demonstrate that the introduction of SmmPack is realistic.
%maintains realistic boot time and BIOS size.

\subsection{Evaluation method}
Both the speed and size of the overhead depend on the number of SMM modules to be packed. The number of SMM modules varies depending on the BIOS. In the case of the BIOS used in this experiment, there were 39 SMM modules. Measurements of clock cycles and size in bytes were calculated using this number of modules; however, the overhead was represented as a percentage.

The boot-time delay caused by SmmPack is the sum of the execution times of SmmPackSmm and the decryption stubs of multiple packed modules.
% The execution time of SmmPackSmm was measured by obtaining the number of clock cycles required from module entry to module return using rdtsc instruction. The execution time of the decryption stub was measured by inserting the rdtsc instruction at the beginning and end of the Unpack function defined in SmmPackSmm. The decryption time depends on the size of the text section of the SMM module being packed.
The execution time was measured by obtaining the number of clock cycles using rdtsc instruction.
Also, the execution time of decryption stubs is proportional to the size of the text section to be decrypted.
The SMM module included in the BIOS used in our experiment had a text section size of approximately 3000 to 50000 bytes, so we measured the number of clock cycles required to decrypt 3000, 20000, and 50000 bytes. All clock cycle counts were calculated by repeating the same work five times and taking the average value. The clock cycles measured are shown in seconds when executed on a 2.0 GHz CPU.

The increase in size was calculated by the sum of the size of SmmPackSmm itself and the sizes of the decryption stubs added to the packed SMM modules. These sizes are fixed, but the total size of the decryption stubs scales linearly with the number of SMM modules being packed.
%To measure precisely, one must also take into account the increase in the number of entries in the PE header and the increase in the number of bytes due to section alignment. However, these are small and vary depending on the section alignment size of the SMM module. Therefore, they are ignored in this measurement.

\subsection{Speed overhead}
The required clock cycle counts and the corresponding execution times on a 2.0GHz CPU for using SmmPack are shown in Table \ref{tab:clocksmm}.
The number of clock cycles required for SmmPackSmm execution was 460702429.0. This is about 0.23 s in a 2.0GHz CPU.
%SmmPackSmm consists of two parts: acquiring a key from the TPM and installing the protocol using the key. The number of clock cycles required to obtain the key in SmmPackSmm was 460638762.5. Therefore, most of the time was spent unsealing the key from the TPM.

In terms of the decryption time, assuming 39 SMM modules, each with a text section size of 20000 bytes, the overhead was approximately 0.13 s. When combined with the execution time of SmmPackSmm, the total overhead in boot time was approximately 0.36 s. This is approximately 2.77\% of the entire boot time in our experimental environment and can be considered a realistic value.

\begin{table}[tb]
\caption{Average clock cycles and time required in SmmPack.}
\label{tab:clocksmm}
\centering
\begin{tabular}{p{43mm}|p{19mm}|c}
\hline
Measurement target & Clock cycles & Time (s) \\
\hline
SmmPackSmm (whole) & 460702429.0 & 0.23035 \\
SmmPackSmm (key retrieval) & 460638762.5 & 0.23032 \\
Unpack 3000 bytes & 1019979.0 & 0.00051 \\
Unpack 20000 bytes & 6766215.4 & 0.00338 \\
Unpack 50000 bytes & 16914954.6 & 0.00846 \\
\hline
\end{tabular}
\end{table}

\subsection{Size overhead}
The size of SmmPackSmm was 11648 bytes, and the size of the decryption stub was 188 bytes. If 39 SMM modules are packed, the total size of the BIOS will increase by 18980 bytes. This was approximately equal to the size of the average UEFI module. This was only 0.1\% of the total data stored in the UP Squared Board Pro's SPI flash chip. Therefore, the increase in size owing to the introduction of SmmPack is negligible and can be considered a realistic value.
\section{Discussion}
\label{sec:discussion}
In this section, we discussed the value of SmmPack as well as the management and adoption, and BIOS updates regarding SmmPack.
%We also discussed TPM vulnerabilities and key acquisitions before the DXE phase, which were outside the scope of our study.

\subsection{Value of SmmPack}
There may be various opinions on whether the obfuscation provided by SmmPack justifies its cost. Furthermore, considering the practical implementation of SmmPack, there would likely be a need for discussions that extend beyond the page count of this paper. However, the true value of this study lies in the fact that it is the first to apply obfuscation to system firmware. We believe that the value of this study lies in shedding light on the effects, costs, the setup of threat models, and identifying critical points of focus when evaluating obfuscation systems towards system firmware.

\subsection{Management and adoption of SmmPack}
To introduce SmmPack, OEM needs to perform certain tasks before and after shipping the BIOS. 
The tasks performed before shipping the BIOS are as follows: (1) determine the key for the BIOS, (2) pack each SMM module, (3) obtain the PCR value when SmmPackSmm is executed, and (4) seal the key into the TPM of the PC that uses the BIOS.
%These operations are all done after the BIOS implementation has been completed.
% The tasks that the OEM should perform before shipping the BIOS are as follows. Note that these operations are all done after the BIOS implementation has completed.
% \begin{enumerate}
%     \item Determine the key for that BIOS.
%     \item Pack each SMM module with the key.
%     \item Obtain the PCR value when SmmPackSmm are executed.
%     \item Seal the key into the TPM of the PC that uses the BIOS.
% \end{enumerate}

There are multiple methods to perform (4). One approach is to boot the OS with the unpacked SMM module included in the BIOS and use tools such as tpm2-tools on top of the OS to extract the key. After sealing, power off the system, replace the BIOS with the BIOS containing packed SMM module, and verify if the system boots properly. Alternatively, the code that seals the key to the TPM can be executed as a UEFI module without booting the OS.
One important consideration is to restrict access to the key solely to the platform firmware entity. Permitting access to the key by the owner entity can lead to the erasure of the key, because of the TPM initialization process executed by the operating system. The task that the OEM should perform after the BIOS is shipped is the BIOS update, as described in the following subsection.
%Notably, whenever the BIOS code content changes as the version is updated, it is necessary to change the key.

Regarding scalability, SmmPack can pack a single SMM module by running a single-packer program. This task can be performed simultaneously with the SMM module build process by incorporating it at the end of the build tool, such as EDK2. Additionally, because SmmPack uses the same key and PCR values for PCs of the same model, it does not require complex key management or provisioning tasks. Moreover, it can be integrated into the existing BIOS update process as explained in the following subsection. Therefore, it can be said that scalability is not compromised.

\subsection{BIOS update}

\begin{figure}[tb]
  \centering
  \includegraphics[width=0.85\linewidth]{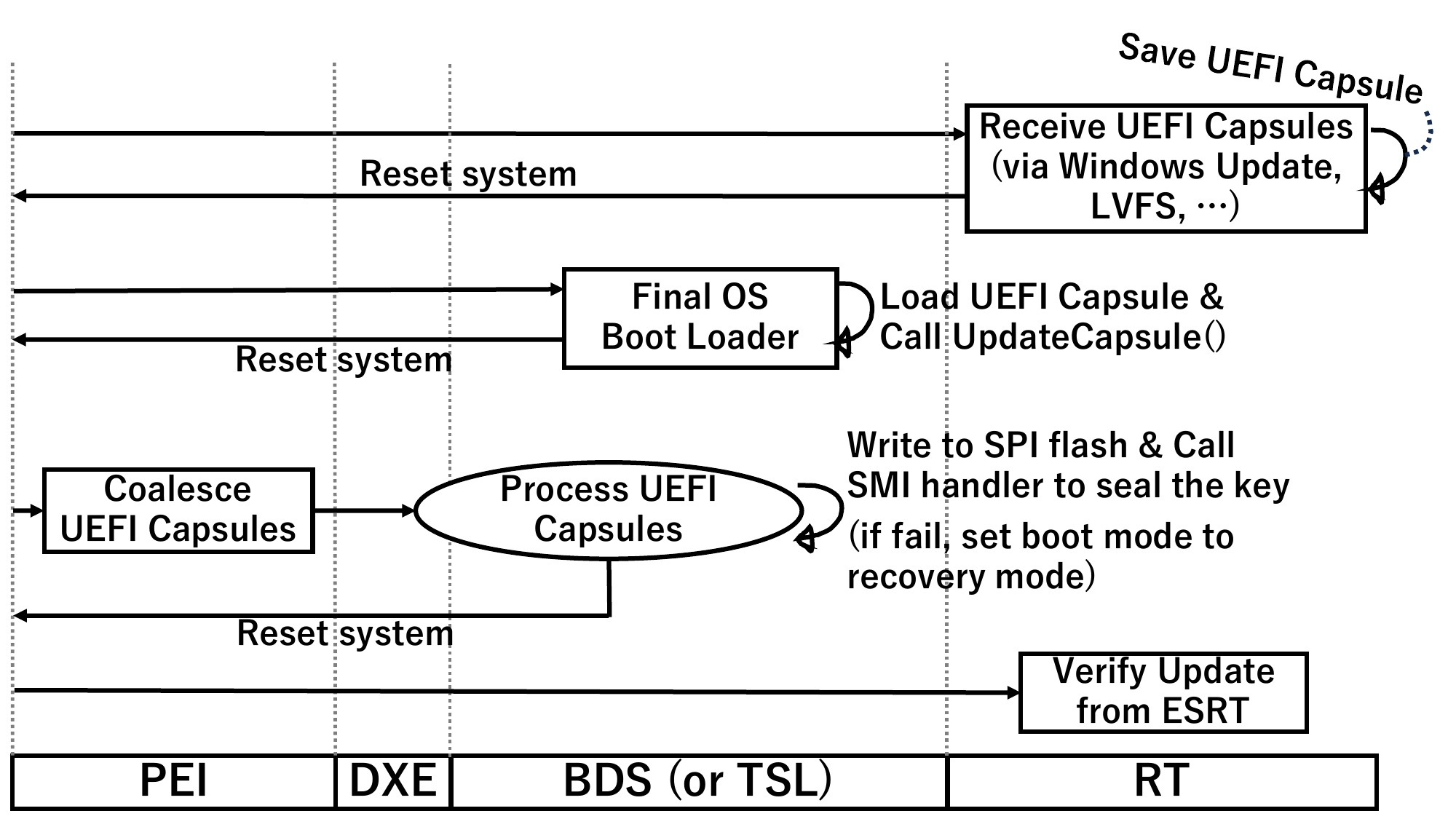}
  \caption{UEFI capsule update with SmmPack (Figure adapted from \cite{fosdemcapsule} for this paper.).}
  \label{fig:capsuleupdate}
\end{figure}

\begin{figure}[tb]
  \centering
  \includegraphics[width=0.85\linewidth]{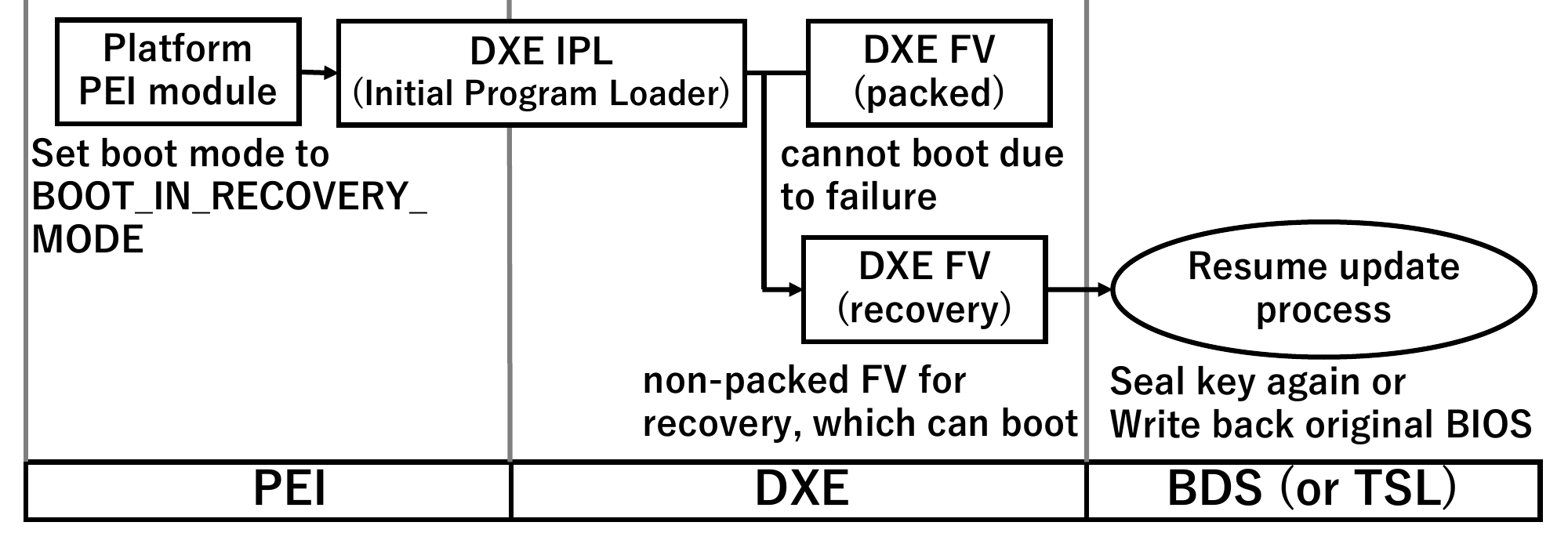}
  \caption{Recovery from update failure.}
  \label{fig:recovery}
\end{figure}

%To truly satisfy the Req 1 defined in Section \ref{sec:requirements}, it is necessary to address the scenario in which SMM modules are obtained from BIOS update files.
To prevent the analysis of the modules obtained from BIOS update file, they must be distributed in a packed state. Moreover, the key and the new PCR value must be stored in the receiver's TPM without being read by the receiver. The key received from the BIOS update server must be sealed in SMM because placing the key in a normal memory region during the update process can be easily read. Therefore, it is desirable to receive the key in an encrypted state and decrypt it inside the SmmPackSmm's SMI handler.
% SmmPackSmm should have an SMI handler that accomplish this task, but communication with the BIOS update server is difficult to implement in SMI handler. Therefore, it is desirable to receive the key in an encrypted state in non-SMM code, pass it to the SMI handler, and decrypt it there.
% TODO: どういう鍵で新しい鍵を暗号化すれば良いかなどはページ数的に述べなくても良い？
%With this method, even if the UEFI module is obtained from a BIOS update file, it cannot be analyzed unless the key is read from SMRAM.
The difficulty in extracting data from SMRAM is demonstrated in Section \ref{sec:analysis}.
%Therefore, it can be concluded that Req 1 was satisfied.

UEFI capsule update\cite{capsule} can still be performed, even with the adoption of SmmPack. Figure \ref{fig:capsuleupdate} illustrates the flow of the update process. The capsule creator first constructs a capsule by including the packed SMM modules, a new encrypted key, and a new PCR value. When the PC receiving the capsule reaches the stage where the coalesce process is done and finishes writing the capsule contents to the SPI flash, it passes the encrypted key and new PCR value to the SMI handler mentioned above to seal the key.

If the key sealing process fails, recovery can still be performed. Figure \ref{fig:recovery} illustrates the flow of the recovery process. 
SmmPack targets SMM modules in the DXE phase, leaving the PEI modules untouched. This allows the system to boot, with or without SmmPack, up to the PEI phase. During the normal recovery process (without SmmPack), when the BIOS update process fails, a specific PEI module retrieves a recovery DXE FV from an external storage device and uses it to boot and perform the remaining update operations \cite{capsule}. With SmmPack, after booting from the recovery image, the sealing process should be performed again at this point. It should be noted that the DXE FV for recovery needs to include an SMI handler that seals the key as described above. If the sealing process still fails, it is possible to write the original BIOS back. In the case of a failure occurring after the original key has been undefined but before sealing the new key, it is recommended to save the original key value in another NV space of the TPM before performing the undefine operation. Once the new key has been confirmed to be successfully sealed, the original key can be deleted. Therefore, the recovery process can also be performed with SmmPack adopted.

\subsection{Shared code problem}
In reality, a non-negligible amount of shared code is found among different vendors \cite{shared1,shared2}. Therefore, even if a company adopts SmmPack, an attacker can analyze the vulnerabilities of the BIOS code of other companies that have not adopted SmmPack and apply the exploit to the former company, if the code they use is shared and this fact is known.
% Furthermore, even within the same company, if SmmPack is applied to one BIOS and not to another that contains shared code, the same attack can be performed.
However, the SMM modules that are most security-conscious are those that will be introduced in the future. When these new SMM modules are introduced in the future, SmmPack obfuscation can be applied to make it significantly more difficult for attackers to analyze their vulnerabilities. In addition, even if a packed SMM module contains shared code, it is not possible to determine whether the shared code is actually present in that module until its contents are analyzed.
\section{Conclusion}
We developed the first obfuscation mechanism in platform firmware, SmmPack. SmmPack is a packing method that uses a key sealed inside the TPM as a security mechanism to increase the cost for attackers who aim to acquire and analyze vulnerabilities in SMM modules. Experimental results using a prototype implementation showed that SmmPack had no adverse effects on the original module, and the increase in boot time and BIOS size was realistic. Furthermore, we clarified the management and adoption methods for SmmPack as well as the procedure for applying BIOS updates.

\subsubsection*{Acknowledgements.}
We would like to express our sincere gratitude to Xeno Kovah for his invaluable expertise and guidance on UEFI security.

\bibliographystyle{splncs04}
\bibliography{refList}
\end{document}